\documentclass[aip,twocolumn,amsmath,amssymb,reprint]{revtex4}
\usepackage{graphicx}
\usepackage{amsmath}
\usepackage{latexsym}
\usepackage{amsmath}
\usepackage{amssymb}
\usepackage{float}
\usepackage{color}
\usepackage{epstopdf}
\usepackage{hyperref}
\hypersetup{urlcolor=red,citecolor=blue}
\usepackage{bm}
\hypersetup{urlcolor=red,citecolor=blue}

\usepackage{epsfig}
\newcommand{\dt}{\Delta t}
\newcommand{\cb}{\color{black}}
\definecolor{g1}{rgb}{0.0,0.0,0.0}
\newcommand{\cg}{\color{g1}}

\begin{document}

\title{{\cb Transferability of self-energy correction in tight-binding basis \\ constructed from first principles}}

\author{Manoar Hossain}
\author{Joydeep Bhattacharjee}
\email{jbhattacharjee@niser.ac.in}
\affiliation{National Institute of Science Education and Research, 
Homi Bhaba National Institute, Jatni-752050, Odisha, India}

\begin{abstract}
We demonstrate in this work the transferability of self-energy(``SE'') correction(``SEC'') of Kohn-Sham(``KS'') single particle states
from smaller to larger systems, {\cb when } mapped {\cg through} localized orbitals constructed from the KS states.
The approach results {\cb in} a SE corrected TB framework, within which,
the mapping of SEC of TB parameters is found to be transferable from smaller to larger systems of similar morphology, 
leading to a computationally inexpensive approach for estimation of SEC in large systems with reasonably high accuracy.
The scheme has been demonstrated in insulating, semiconducting and magnetic nanoribbons of graphene and 
hexagonal boron nitride, where SEC {\cb tends to strengthen} the individual $\pi$ bonds, leading to
transfer of charge from edge to bulk.
Additionally in magnetic bipartite systems SEC tends to enhance inter-sublattice spin separation.
The proposed scheme thus promises to enable estimation of SEC of band-gaps of large systems without needing
to {\cg explicitly} calculate SEC  of KS single particle levels which can be computationally prohibitively expensive.
\end{abstract}

\maketitle

\section{Introduction}
{\cb Designing} new materials at nanoscale, typically consisting of few tens of atoms, 
{\cb necessitates}  increase in accuracy of estimation of electronic structure 
preferably without a commensurate increase in computational cost.
Particularly with increasing spatio-temporal resolution of synthesis\cite{nanosyn1}, spectroscopic 
\cite{nanospectra1}, and transport \cite{nanotrans3} 
measurements of nanostructures,
it has become imperative to match measured values to computed results 
in order to precisely determine {\cb atomic constitution of samples}. 
Accordingly, computational methodologies have been evolved \cite{tb_ordern1}
over the years for estimation of electronic structure of systems typically
with a few hundreds of electrons, large enough to be  within the experimentally accessible length-scales,
primarily at the level of Kohn-Sham (KS) density-functional theory (DFT)\cite{kohn1965self} \cite{hohenberg1964inhomogeneous}.
As a possible approach to compute self-energy corrected energetics of electrons in such large systems,
in this work 
{\cb we demonstrate bottom-up transferability of self-energy correction when
incorporated in a suitable tight-binding basis constituted from first principles.} 

Mean-field approximation of the Kohn-Sham (``KS'') density-functional theory (``DFT'') 
\cite{kohn1965self} \cite{hohenberg1964inhomogeneous},
has established itself as a powerful tool for calculation of electronic structures of materials 
from first principles, to study ground state properties with reasonable accuracy, primarily in systems 
with weak localization of electrons in the valence sub-shells.
Wannier functions  \cite{wannier1937structure,marzari1997maximally,jbwan},
constructed from KS single particle states, have been used as 
TB basis\cite{atomicorb1,atomicorb2,atomicorb3,calzolari2004ab,franchini2012maximally,jung2013tight}
to derive model Hamiltonians to focus only on the relevant group of orbitals.
{\cb However, DFT being essentially a ground state theory, the inherent lack of discontinuity \cite{derivdiscont1}
of the derivative of the static and  local \cite{lda1} or semi-local \cite{gga2} 
mean-field approximations of exchange-correlation functionals, upon addition or removal of electrons, 
leads to underestimation of band-gap compared to
their experimentally measured values as the difference between the ionization potential (``IP'') and electron affinity (``EA'').}
{\cb As a result}, TB parameters computed from DFT often need further tuning parameters to match experimental data, 
such as band-gap, particularly in systems with increased correlation mainly due to localized electrons \cite{cococcioni2012lda+} 
Multitude of efforts to address these inadequacies have been pursued over {\cb last few decades}, 
within and beyond the framework of DFT.
Improvement of the exchange-correlation functionals 
either by correcting for derivative discontinuity explicitly \cite{derivdiscont2,derivdiscont3} , 
or more popularly through incorporation of the inherently non-local nature of many-electron interactions 
by deriving non-local functionals\cite{metagga} and partial inclusion of 
Hartree-Fock exact exchange 
in hybrid functionals \cite{heyd2003hybrid},\cite{becke1993new}
have been reasonably successful in addressing particularly the issue of underestimation of band-gap by DFT, 
with appropriate choice of relevant parameters.

A more general parameter free approach beyond the framework of DFT, is the many-body perturbation
 theory(``MBPT'') \cite{onida2002electronic} \cite{louie2006conceptual}
wherein, many-electron effects are treated as perturbation, resulting in description of interacting electrons 
as quasi-particles(QP) whose energies include corrections to the KS single-particle levels due to effective holes 
associated with electrons in lieu of their interaction with other electrons.
These corrections, thereby known as self-energy (``SE'') corrections(``SEC''), computed
up to the first order, have been shown\cite{aryasetiawan1998gw}
to be sufficient  in accounting for the experimentally accessible SE corrected band-gap (IP-EA). 
However, since both the approaches -  hybrid functionals and MBPT, are computationally expensive, 
MBPT being more so, {\cb scaling typically as $N^4$ with system size},
it poses a formidable computational challenge to compute SE corrected band-gap till date even for nanostructures 
with dimensions in single digits of nanometers, using standard computational platforms. 
Notably though, considerable amount of effort and progress has been made in recent years in {\cb reducing}
the computational cost by using specialized basis sets\cite{gworb1,gworb2,gworb3,gwordern}.

In this work our approach has been to first incorporate the SEC of KS single particle levels as corrections to
TB parameters in a suitable basis  
and subsequently see if such corrections
derived from a smaller reference systems can be reasonably applied to TB parameters derived for larger systems
for realistic representation of SEC in such systems without needing to explicitly compute SEC for KS states 
which can be computationally prohibitively expensive with growing system size. 
{\cb Estimation of} QP band-gap within a TB framework has been attempted in {\cb recent} years\cite{qptb7,qptb6,qptb5,qptb4}, 
largely based on tuning TB parameters to match the relevant QP structure.
{\cg In the following we first discuss construction of the atomic orbital basis from the KS single-particle states, 
 in which SEC is mapped,
followed by brief description of the {\it GW} approximation of MBPT used in this work for calculation of SEC.}
We demonstrate our approach in graphene nano-ribbons(GNR), where {\it GW} approximation of  SEC  
has been reported in details \cite{yang2007quasiparticle}, 
and also in  hexagonal boron-nitride ribbons (hBNNR) as example of wide band-gap insulator.

\section{Methodological details}

The TB basis used in this work are orthonormal Wannier\cite{wannier1937structure,marzari1997maximally,jbwan}
 orbitals constructed as linear combination of KS energy eigen-states
with a specific choice of gauge that maximally retains their character as individual atomic orbtial.  
{\cg
Since in this work we consider only 2$p_z$ orbitals, we limit our discussion here on generation of 
one orbital per atom. We begin with a template consisting of one of the 2$p$ orbitals each of B, C, N
 calculated using norm-conserving pseudo-potentials.
The 2$p$ orbital is chosen to be arbitrarily one of the lowest three degenerate set of KS states of an 
isolated B, C or N atom.  
The full system is then decorated with such orbitals to associate  one single
$2p_z$ orbital with each atom, aligned perpendicularly to the local plane defined by the nearest neighbourhood.
These orbitals constitute a non-orthogonal set of localized basis from which 
a set of quasi-Bloch states are constructed as:  
\begin{equation}
\tilde{\psi}_{\vec{k},j}(\vec{r}) = \frac{1}{\sqrt{N}} \sum_{\vec{R}} e^{i \vec{k} \cdot \vec{R}} \phi_{\vec{R},j} (\vec{r}),
\end{equation}
where $ \phi_{\vec{R},j} (\vec{r})$ is the $j$-th member of the non-orthogonal basis localized
in the unit-cell denoted by the lattice vector $\vec{R}$ which spans over $N$ unit-cells that define the 
periodicity of the Bloch states.
Next we calculate the projection of the non-orthogonal quasi-Bloch states on the orthonormal 
Bloch states constructed from the cell-periodic KS-single particle states at all allowed $\vec{k}$, as:
\begin{equation}
 O_{\vec{k},m,j}= \langle \psi^{KS}_{\vec{k},m}  \mid \tilde{\psi}_{\vec{k},j} \rangle.
\label{proj}
\end{equation}   
Subsequently the overlaps between the representation of the non-orthogonal quasi-Bloch states within the manifold of 
of the KS single-particle states, are calculated as:
\begin{equation}
S_{\vec{k},m,n}= \sum_l O^*_{\vec{k},m,l} O_{\vec{k},n,l}. 
\label{overlap}
\end{equation}   
Finally, we use the L\"{o}wdin symmetric orthogonalization \cite{lowdin1950non} scheme to construct a new set of 
orthonormal Bloch states from the KS single particle states as:
\begin{equation}
\Psi_{\vec{k},n}(\vec{r}) = \sum_m S^{-\frac{1}{2}}_{\vec{k},m,n} \sum_l O_{\vec{k},l,m} \psi^{KS}_{\vec{k},l}(\vec{r}),
\label{awobf}
\end{equation}
using which,  a set of localized orthonormal Wannier functions  are constructed as:
\begin{equation}
\Phi_{\vec{R'},j}(\vec{r}) = \frac{1}{\sqrt{N}} \sum_{\vec{k}}  e^{-i \vec{k}\cdot \vec{R'}} \Psi_{\vec{k},j}(\vec{r}).
\label{awo}
\end{equation}   
L\"{o}wdin symmetric orthogonalization thus provides a choice of gauge for linear combination of KS states such that the
resultant Wannier functions $\left\{ \Phi_{\vec{R'},j}(\vec{r}) \right\}$ would have minimal deviation from 
the non-orthogonal orbtials  $\left\{ \phi_{\vec{R},j} (\vec{r})  \right\}$. Hence we here onwards refer these 
Wannier functions as atomic Wannier orbtials ("AWO"). }
TB parameters are computed in the AWO basis as:
\begin{eqnarray}
& &t_{\vec{R'},\vec{R},i,j} = \langle \Phi_{\vec{R'},i} \mid H^{KS} \mid \Phi_{\vec{R},j} \rangle \nonumber \\
&=&\sum_{\vec{k}}e^{i\vec{k}.(\vec{R'}-\vec{R})}\sum_l (OS^{-\frac{1}{2}})^*_{li} (OS^{-\frac{1}{2}})_{lj} E^{KS}_{\vec{k},l}
\label{hop}
\end{eqnarray}
As obvious, the AWOs used here can in principle be substituted by any localized atomic orbitals constructed as linear combination
of KS single particle states.  
\begin{figure*}[t]
\includegraphics[scale=0.55]{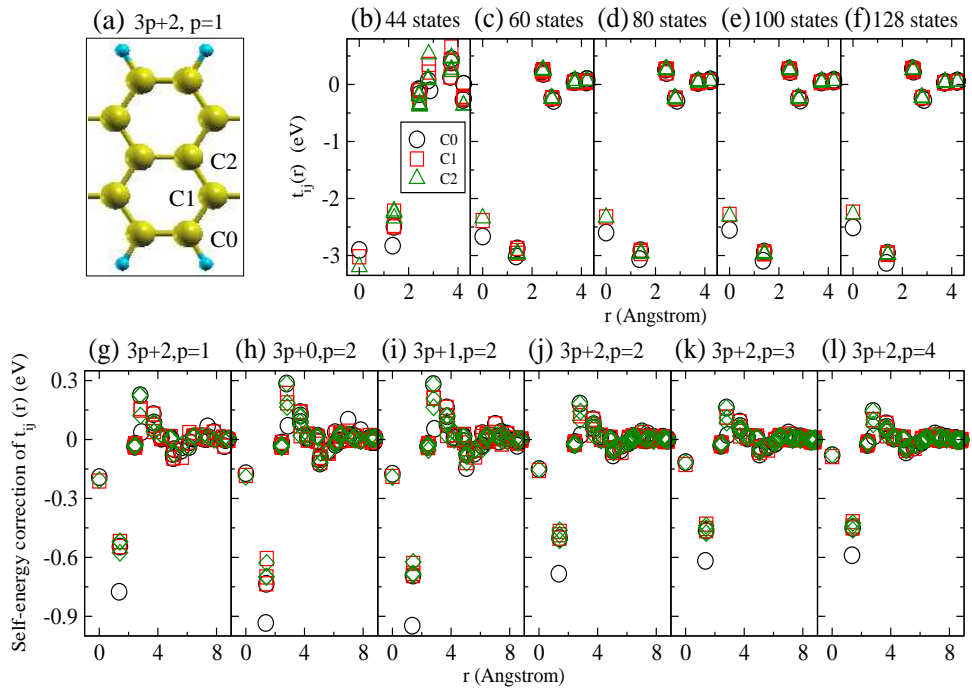}
\caption{For the three inequivalent atoms marked in (a):
(b-f): convergence of TB parameters ($t_{ij}(r)$)  in terms of the number of KS bands
considered for construction AWO(2$p_z$); (g-l): self energy correction  to 
TB parameters ($\Delta t_{ij}(r)$) in AGNRs of different families of increasing width.}
\label{agnrtijconv}
\end{figure*}


To estimate self-energy("SE") correction("SEC") of the KS single particle levels we followed the
{\it GW} approximation\cite{hybertsen1986electron} of Hedin's formulation\cite{hedin1965new} \cite{hedin1970effects} of the  
many-body perturbation theory ("MBPT") \cite{onida2002electronic} \cite{louie2006conceptual}
 to describe single-particle excitations,
wherein,  many electron interactions
are represented beyond mean-field by an energy dependent, non-local and non-Hermitian SE 
operator $\Sigma$, derived by considering the many-electron effects as perturbation treated 
within a self-consistent framework of Dyson's equation in terms of the one-particle non-local Green's functions $G$, as:
\begin{equation}
\Sigma(\vec{r},\vec{r}',E)=\frac{i}{2\pi}\int dE' e^{-i\delta E'} G(\vec{r},\vec{r}',E-E') W(\vec{r},\vec{r}',E'),\nonumber
\label{sigma}
\end{equation}
where $\delta = 0^+$ and 
$W$ is the Coulomb interaction 
screened by a non-local dynamic dielectric screening function computed approximately  within the random phase approximation 
as extension of its static counterpart 
to finite frequencies following a generalized plasmon pole model\cite{hybertsen1986electron}.
With the underlying assumption that correction to the KS single particle states are negligible,
the quasiparticle energies are approximated as:
\begin{equation}
E^{QP}_{\vec{k},n}=E^{KS}_{\vec{k},n}+\langle \psi^{KS}_{\vec{k},n} \mid \Sigma - V^{KS}_{xc} \mid  \psi^{KS}_{\vec{k},n} \rangle
\label{equasi}
\end{equation}
where $ V^{KS}_{xc}$ 
is the mean-field exchange-correlation potential derived from the exchange-correlation functionals used in DFT.
Estimation of quasi-particle energies within {\it GW} approximation is computationally expensive primarily 
due to the slow convergence of $\epsilon^{-1}$ and $G$, and therefore of 
$\Sigma$, with respect to unoccupied single particle KS states. 
%
Substituting $E^{KS}_{\vec{k},n}$ in Eqn.(\ref{hop}) by quasiparticle energies $E^{QP}_{\vec{k},n}$ we calculate
the SE corrected TB parameters $\left\{t^{QP}_{\vec{R'},\vec{R},i,j} \right\}$.
SEC of the TB parameters is thus estimated as: 
\begin{equation}
\Delta t_{\vec{R'},\vec{R},i,j} = t^{QP}_{\vec{R'},\vec{R},i,j} - t^{KS}_{\vec{R'},\vec{R},i,j},
\label{sectb}
\end{equation}

\section{Computational Details}

There are two main steps in the proposed approach: (1) computation of the ground state electronic structure
followed by construction of AWOs as per Eqn.(\ref{awo}) using the KS single particle states transformed as shown in Eqn.(\ref{awobf})
followed further by calculation of TB parameters as per Eqn.(\ref{hop}), 
(2) calculation of SEC of KS single particle states using {\it GW} approximation and subsequent estimation of SEC of TB parameters 
[$\Delta t$] from QP energies. 

Ground state electronic structures are calculated using a plane-wave based implementation of
DFT \cite{giannozzi2009quantum}.
Unit cells are structurally optimized with variable cell size. 
Ground state energies are calculated using norm conserving
pseudo-potentials with Perdew-Zunger (LDA) exchange-correlation \cite{lda1}
functional and plane wave energy cutoff of 60 {\cb Rydberg}. 
Grid of $\vec{k}$-points 1x1x15 and 1x1x29 are used for AGNRs and ZGNRs respectively.
Separation of more than 10 Angstrom is used between periodic images of nano-ribbons. 
We {\cg restrict} to the non-self-consistent ($G_0W_0$) level for estimation of quasi-particle energies using the
BerkeleyGW (BGW) implementation \cite{deslippe2012berkeleygw}.
Parameters for calculation of SEC have been chosen as per Ref.33. 
Band-gaps have been further converged with respect to finer $\vec{k}$ grid through interpolation based on the AWOs.
Construction of AWO, calculation of TB parameters in AWO basis, and estimation of SEC of TB parameters are performed
using our implementation interfaced with the {\cg Quantum} Espresso code.
\begin{figure*}[t]
\includegraphics[scale=0.4]{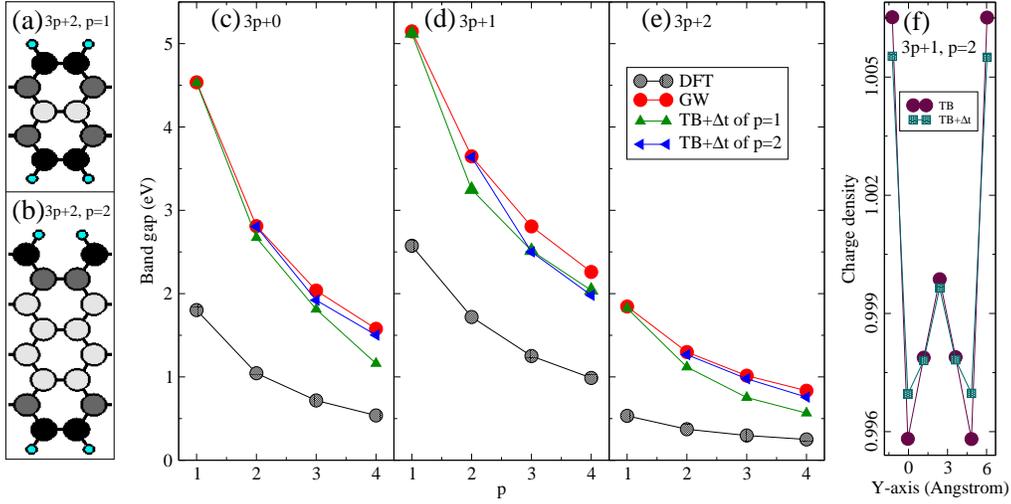}
\caption{With transfer of ($\dt$) exemplified in (a) and (b):  (c-e): comparison of band-gaps rendered by
DFT,DFT+$G_0W_0$, and TB+$\dt$(of p=1 \& 2) for three families of AGNRs with increasing width; (f) charge density from TB and TB+$\dt$.}
\label{agnrbandgap}
\end{figure*}
\begin{figure}[b]
\includegraphics[scale=0.24]{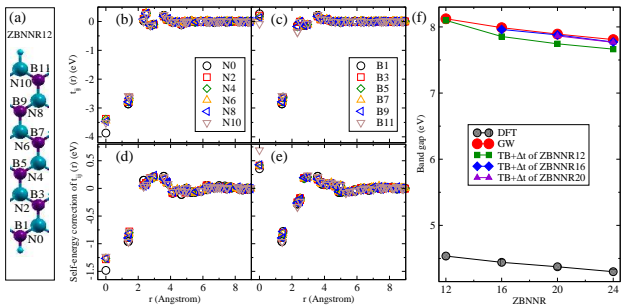}
\caption{For the inequivalent atoms marked in (a):
(b-c): TB parameters ($t_{ij}(r)$) in ZBNNR12; (d-e): corresponding self energy correction 
$\Delta t_{ij}(r)$ ; (f) comparison of band-gaps rendered by
DFT, DFT+$G_0W_0$, and TB+$\dt$(of ZBNNR12) for wider ZBNNRs. }
\label{zbnnr}
\end{figure}

\section{Results and Discussion}
In the following we primarily demonstrate our approach by accounting for SEC of KS band-gap in wider ribbons through correction to 
their TB band-gap calculated using SEC of TB parameters ($\Delta t$) obtained in a narrower ribbon.
We chose AGNRs, ZGNRs and hBNNRs in order to span a wide range of band-gaps and magnetism with only 2$p_z$ electrons.
\begin{figure*}[t]
\includegraphics[scale=0.5]{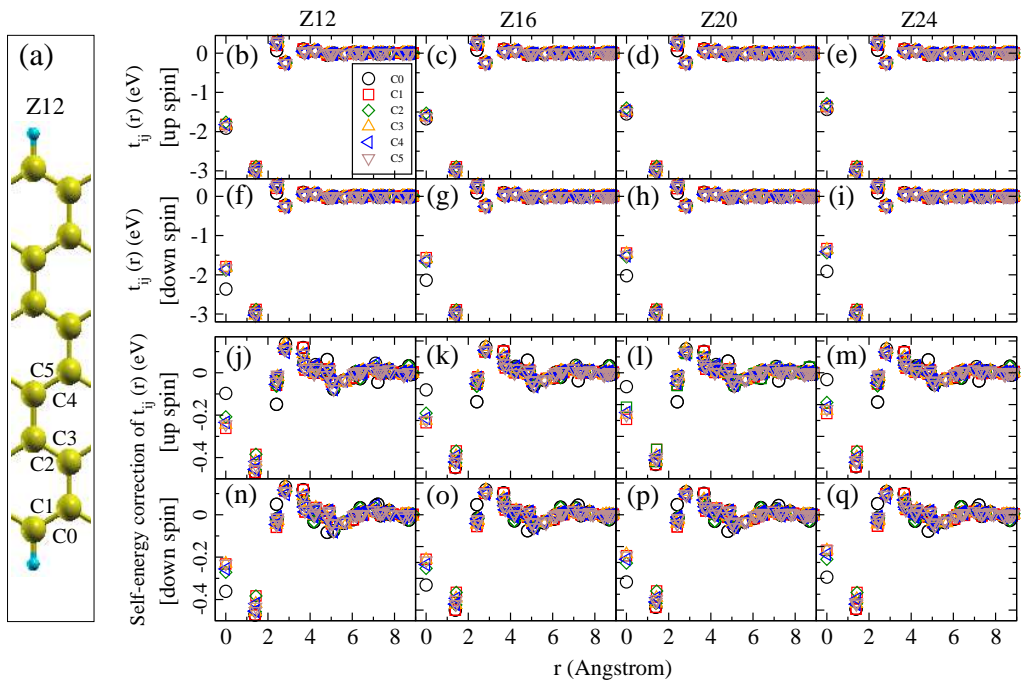}
\caption{For the inequivalent atoms marked in (a) from edge towards bulk:
TB parameters ($t_{ij}(r)$) for spin-1(b-e) and spin-2(f-i) in ZGNRs of increasing width; 
corresponding self energy corrections $\Delta t_{ij}(r)$ 
for spin-1(j-m) and spin-2(n-q).}
\label{zgnrtij}
\end{figure*}

\subsection{Armchair Graphene Nanoribbons}
We consider H-passivated AGNRs and ZGNRs of varying width
wherein AGNRs are specified by a number of the dimer lines and ZGNRs by the number of zigzag chains.
 AGNRs are categorized in three different families as per the number($n$) 
of dimer lines: $n=3p+0$, $n=3p+1$ and $n=3p+2$, $p$ being an integer.
In this work we consider only the 2$p_z$  orbitals since they are well known to adequately describe the edge of the 
valence and conduction bands in GNRs. 
To maximize the 2$p_z$ character  of the corresponding  AWOs, sufficient participation of the anti-bonding orbitals
are required in order to match the weightage of bonding orbitals, which in GNRs are represented  increasingly by bands 
closer to the conduction and valence band edges respectively.
Fig.\ref{agnrtijconv}(b-f) accordingly shows convergence of $t_{ij}$ for all inequivalent atoms with respect to the 
total number of KS bands considered for construction of the AWOs.
The fluctuation of hopping parameters from nearest to next-nearest and beyond, represent 
the favorable(unfavorable) nature of hopping between dissimilar(similar) sub-lattices, 
as generic in bipartite systems.
As per Fig.\ref{agnrtijconv}(d), we consider {\cb on the} average about extra 40 unoccupied states from the valence band
for construction of AWOs. 

Fig.\ref{agnrtijconv}(g-l) shows SEC of  TB parameters $\left\{\Delta t_{ij}\right\}$ for all the inequivalent atoms
 to their neighbours, for 
a representative set of AGNRs from all the three families, arranged in increasing order of width.
Fig.\ref{agnrtijconv}(g,j,k,l) suggests that  correction to nearest neighbor(n-n) hopping reduced marginally after p=1
and convergence beyond p=2. 
However, Fig.\ref{agnrtijconv}(g-i or h-j) indicates that corrections for 3p+1 and 3p+0 are lower than
that of 3p+2 for same p, consistent with the fact that AGNRs with n=3p+2 
 are inherently metallic in nature with a small gap arising exclusively due to variation in TB parameters
 from the edge to the bulk due to relaxation of bond lengths.
Notably, the n-n hopping term, which is between  dissimilar sub-lattices, 
has the most significant negative correction implying consolidation of the 
n-n $\pi$-bond leading to enhanced localization of the $\pi$-bonding orbitals between atoms due to SEC.
Positive correction of further hopping term between dissimilar sub-lattices also imply the same.
Such localization all across the system, as implied by  similar correction to hopping between nearest sites for all inequivalent atoms,
would in effect result into withdrawal of charge from edge towards bulk, as evident Fig.\ref{agnrtijconv}(f),  
due to consolidation of $\pi$ bonds in the bulk.
The resultant overall increase in uniformity of charge distribution effectively reduces mutual Couloumb repulsion
between electrons of opposite spins, leading to lowering of the on-site term due to SEC.
Consistent lowering of correction to the on-site term with increasing width indicates reduced levels of SEC in 
general with increasing value of p.  
Correction to the hopping between sites within the same sub-lattice, like the hopping between next nearest sites, 
 is negligible since it is weak in the DFT level itself.

TB Band-gap with TB parameters in AWO basis derived from KS eigen-states of a given system,  
would match the KS band-gap of the same system by construction. 
Similarly, with  SEC of TB parameters, referred here onwards as $\dt$, 
the correction to TB band-gap (SEC(TB)),  would match the SEC of KS band-gap (SEC(KS)) of a given system,
if  $\dt$ is obtained from SEC of KS single particle levels of the same system,
as evident in Fig.\ref{agnrbandgap}(c-e) for p=1 and p=2. 
Motivated by the overall similarity in SEC of TB parameters within {\cb each families} of AGNRs shown in Fig.\ref{agnrtijconv}(g,j,k,l),
we next test if SEC(KS) of a wider AGNR with p $>$ 1 can be matched by SEC(TB) estimated with TB parameters calculated from KS states of the 
same system (p $>$ 1), but using $\dt$ obtained for p=1 ($\dt(p=1)$) of the same family. 
As evident in Fig.\ref{agnrbandgap}(c-e), within each family, using $\dt(p=1)$ and $\dt(p=2)$,   
it is possible to account for more than 80\% of SEC(KS) in wider AGNRs ($p=3,4$), 
with no appreciable increase in computational cost beyond computation of TB parameters for wider ribbons. 
Owing to the convergence of $\dt$ beyond p=2 [Fig.\ref{agnrtijconv}(g,j,k,l)], 
the match between SEC(TB) and SEC(KS) is more accurate with $\dt(p=2)$  than with $\dt(p=1)$. 
The scheme for assignment of $\dt$ from narrower to wider AGNRs is shown in Fig.\ref{agnrbandgap}(a-b) 
where the atoms of matching colors are assigned same corrections.
{\cg
Correction data is collected for each single atom of the reference system for all
its neighbouring pairs within a cutoff radius chosen to include typically 
up to 4th or 5th nearest neighbour beyond which the corrections are practically negligible.
The transfer is done on the basis of two considerations: 
(1) matching atoms between the reference and the target systems  in terms of their neighbourhood not limited to 
nearest neighbours, 
and (2)  by maximally matching distance between pair of atoms in the reference system to that the target system.
In mapping atoms for the criteria (1) the similarity of average nearest-neighbour bond-lengths around atoms 
can be used as a reasonable criteria.
}

\subsection{hBN nanoribbons}
Next we demonstrate the scheme in hBNNR, chosen as an example of wide band gap insulator where the SEC(KS) is substantial.  
The difference of electro-negativities of B and N are reflected in the difference in on-site terms in Fig.\ref{zbnnr}(b,c).
$\dt$ plotted in Fig.\ref{zbnnr}(d,e) show mild positive and strong negative SEC for on-site terms for B and N respectively, implying
enhanced polarity of the B-N $\pi$-bond and consolidation of the lone pair of N.
Noticeably, unlike in GNRs, $\dt$ in all B and N atoms are very similar among their own kind irrespective of their proximity to edges, 
except the ones exactly at the edges.
This is expected to enhance the degree of transferability of $\dt$ across hBN ribbons systems.
Fig.\ref{zbnnr}(f) indeed suggests SEC(TB) to cover more than 90\% of SEC(KS) in wider ZBNNRs with  
$\dt$ calculated in the  narrowest of the hBNNRs considered.
\begin{figure}[b]
\includegraphics[scale=0.33]{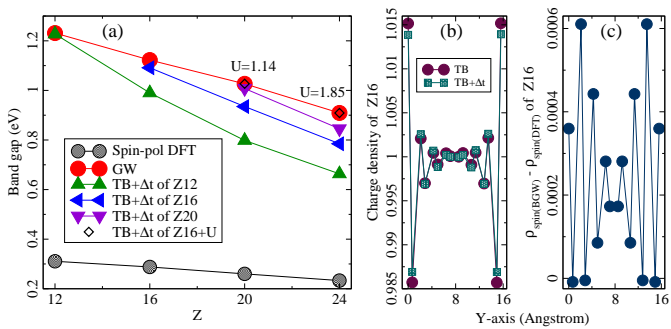}
\caption{(a) Comparison of band-gaps rendered by DFT,DFT+$G_0W_0$, and TB+$\dt$(of Z12 \& Z16) for ZGNRs of larger width (Z20,Z24); 
(b) charge density from TB and TB+$\dt$; (c) difference of absolute values of spin-densities($\rho_{spin}=n_{\sigma}-n_{\sigma'})$). }
\label{zgnrbandgap}
\end{figure}

\subsection{Zigzag edged graphene nano-ribbon}

We next demonstrate the scheme in ZGNRs as an example of narrow band-gap magnetic materials where Coulomb
correlation plays a central role in determining the electronic structure. 
The difference in on-site energies [Fig.\ref{zgnrtij}(b-i)] of the two spins at the zigzag edges, 
owes to  spin separation between the two sub-lattices, which leads to localization of 2$p_z$ electrons 
of opposite spins at the two edges characteristic of ZGNRs. 
{\cg
Accordingly, although the C atoms in AGNRs and ZGNRs have the same local neighbourhood, their
TB parameters and their SEC are expected to be  fundamentally different since such
spin-separation is completely absent in the former.
}
Notably, for the C atoms at the ZGNR edges, the SEC of the on-site terms Fig.\ref{zgnrtij}(j-m) shows higher 
negative correction for the local majority spins, compared to those of the C atoms at the interior.
This implies enhanced presence of 2$p_z$ electron of one of the spins at an edge and removal
of electron of the other spin from that edge, as a result of SEC. 
This enhancement in spin separation across the width of ZGNRs is evident in Fig.\ref{zgnrbandgap}(c),
while Fig.\ref{zgnrbandgap}(b) implies withdrawal of charge from edge to bulk, as seen in AGNRs as well, due to SEC.
Notably, while variation in n-n hopping itself [Fig.\ref{zgnrtij}(b-i)] is small among all inequivalent C atoms
in each ZGNR and similar for all ZGNRs, the magnitude of $\dt$ for n-n hopping reduces from Z12 to Z16 and 
converged thereafter[Fig.\ref{zgnrtij}(j-q)] for both spins.
Accordingly, Fig.\ref{zgnrbandgap}(b) suggests a better accounting of SEC(KS) of wider ZGNRs(Z20,Z24) 
using $\dt$ of Z16 than that using $\dt$ of Z12.

Pertinently, transfer of charge from edge to bulk as seen in Fig.\ref{zgnrbandgap}(b), accompanied by enhancement of 
localization of opposite spins near the edges seen in Fig.\ref{zgnrbandgap}(c), is also observed within the Hubbard model with increasing
strength of the on-site Coulomb repulsion U, although it is clear from the contribution of off-diagonal terms of
$\dt$ in reproducing SEC(KS), that DFT+U alone will not be sufficient to account for SEC.
However, unlike AGNRs, ZGNRs being magnetic systems, it is reasonable to anticipate that  
SEC of KS single particle states will also have impact on the on-site Coulomb repulsion term U in addition to TB parameters.
Therefore while substituting $\dt$ of wider ZGNRs by that of Z12 or Z16, we need to account for a possible
underestimation of U. 
We therefore take recourse to the mean field approximation of the Hubbard model and self-consistently introduce $\Delta U$
along with $\dt$, as
\begin{equation}
 H = \sum_{i,j,\sigma}(t_{ij}+\Delta t_{ij}) c^\dagger_{i\sigma}  
 c_{j\sigma}+\sum_{i,\sigma} \Delta U n_{i\sigma}\langle n_{i\sigma^\prime}\rangle,  
\end{equation}
where $\left\{ t_{ij} \right\}$ are computed from KS eigen-states of the wider ZGNRs(Z20,Z24), and  
$\left\{\Delta t_{ij} \right\}$ from Z16.
$\Delta U$ is tuned to match SEC(TB) to SEC(KS) in Z20 and Z24. 
As evident in Fig.\ref{zgnrbandgap}(a), indeed with application of a small U in addition to 
the $\dt(Z16)$, it is possible to match the  SE corrected KS band-gap of the wider ribbons with
modest increase in spin density near the edges, implying enhanced inter-sublattice spin separation 
due to SEC, which is already hinted in Fig.\ref{zgnrbandgap}(c) which is with U=0.

\section{Conclusions}
In conclusion, we have presented a computationally inexpensive scheme for estimation of  self-energy correction(SEC) of band-gap 
within a tight-binding(TB) framework in the basis of atomic Wannier orbitals (AWO) constructed from KS energy eigen-states. 
Within the scheme, SEC of TB parameters are first computed for a smaller reference system from SEC of 
KS single particle levels estimated using the {\it GW} approximation of MBPT, and then applied to 
TB parameters derived for a larger system of similar morphology, in order to estimate SEC of KS band-gap of the larger system,
without needing to explicitly compute it.  
The efficacy of the approach, demonstrated in semiconducting and insulating as well as magnetic and non-magnetic nano-ribbons 
of graphene and hexagonal boron-nitride, is found to account for about 90\% or more of the SE corrected band-gap for 50\% to 100\% increase
in system size as assessed in this work, with nominal increase in computational cost.  
{\cg
Notably, the degree of agreement[Fig.(\ref{agnrbandgap})(c-e),Fig.(\ref{zbnnr})(f)] 
between band-gaps estimated with mapped SEC in TB basis, 
and those directly computed (DFT+$G_0W_0$), clearly suggests that the scope of transferability should 
easily cover further increase in system size compared to that of the reference systems, 
particularly with increasing band gap, whereas, the only major computation beyond DFT for the larger systems
in our approach is the computation of overlap matrices required for L\"{o}wding symmetrization and calculation of TB parameters,
 which should scale as $\mathcal{O}(N^2)$, compared to overall $\mathcal{O}(N^4)$ scaling of $GW$ approximation.
}
In general, the SEC corrected TB framework opens the scope for in-depth analysis of SEC without having to explicitly  generate
SE corrected KS states. The results presented here pave the way for building up a repository of self-energy corrected TB parameters of 
different atoms at different chemical environment for their seamless use in estimation of SEC within a multi-orbital TB framework.

The  data  that  support  the  findings  of  this  study  are  available from the corresponding author upon reasonable request.



%

\end{document}